\documentclass[page-classic]{epl2} 

\title{Dzyaloshinskii-Moriya interaction and magnetic \\ordering in 1D 
  and 2D at nonzero $T$}

\author{F. Torres\inst{1,2} \and D. Altbir\inst{2,3} \and M. Kiwi\inst{1,2}}
\shortauthor{F. Torres \etal}

\institute{ \inst{1} Departamento de F{\'\i}sica, Facultad de
  Ciencias,
  Universidad de Chile, Casilla 653, Santiago, Chile 7800024\\
  \inst{2} Centro para el Desarrollo de la Nanociencia y la
  Nanotecnolog\'\i a, CEDENNA, Avda. Ecuador 3493, Santiago, Chile
  9170124\\
  \inst{3} Departamento de F\'{\i}sica, Facultad de Ciencia,
  Universidad de Santiago de Chile, Casilla 307, Santiago, Chile }

\abstract{The inclusion of a Dzyaloshinskii-Moriya short range
  antisymmetric interaction induces spontaneous magnetization, at
  nonzero temperatures, in one and two dimensions. It is shown that
  quantum fluctuations are reduced by the Dzyaloshinskii-Moriya
  interaction, but short range correlations are increased, thereby
  allowing to obtain long range magnetic order in these low
  dimensional systems.}

\begin{document}

\maketitle

\section{Introduction}
In 1966 Mermin and Wagner~\cite{MerminWagner66} proved a most relevant
theorem, namely that ``For one- or two-dimensional Heisenberg systems
with isotropic interactions, and such that the interactions are short
ranged, namely, which satisfy the condition
\begin{equation}
  \sum_{\bf R} {\bf R}^2 |J({\bf R})| < +\infty \ ,
\end{equation}
there can be no spontaneous ferro- or antiferromagnetic long-range
order at $T>0$.'' With only very few rigorous results available this
theorem constitutes a most valuable piece of knowledge, especially to
test the validity of the usual approximate results. The validity of
the theorem was extended, also using the Bogoliubov
inequality~\cite{Bogoliubov62}, to classical interacting particles by
Mermin~\cite{Mermin68}, and to fermion and boson systems by
Hohenberg~\cite{Hohenberg67}.

In 2001 Bruno~\cite{Bruno01} extended these results even further, to
long range RKKY interactions. More precisely, as formulated by Bruno
as a corollary, ``A $D$-dimensional ($D$ = 1 or 2) Heisenberg or XY
system with interactions monotonically decaying as $|J({\bf R})|
\propto R^{-\alpha}$, with $\alpha \ge 2D$, cannot be ferro- or
antiferromagnetic.''  Physically, it is the fluctuations that prevent
the onset of long range order competing against the correlations
induced by short range interactions. Therefore, the long range
magnetic order observed in one or two dimensional systems could be
due, for instance, to magnetic anisotropies or external magnetic
fields. In this paper we present an alternative approach, where
spontaneous ordering of low dimensional magnetic systems is due to the
symmetry breaking that the short range Dzyaloshinskii-Moriya (DM)
interaction~\cite{Dzyaloshinsky58} generates, and which to the best of
our knowledge has not been reported in the literature.

The physical basis for the Mermin-Wagner (MW) theorem seems to be the
existence of degrees of freedom that are not constrained by an
interaction, which makes the fluctuations strong enough to prevent
long range order. However, in spite of the fact that the MW theorem
excludes the possibility of ordering for a wide range of finite range
interactions, we prove below that the Dzyaloshinskii-Moriya (DM)
interaction for the Heisenberg Hamiltonian, in spite of
  being of short range, leads to spontaneous magnetic order in one
and two dimensions, at non-zero temperatures. In fact the DM
interaction, by reducing the spin fluctuations, yields a canted spin
arrangement which turns out to be stable in one and two
dimensions. 

\section{The Hamiltonian}
The Heisenberg Hamiltonian $H_0$, including a weak Zeeman term, is
given by
\begin{equation}
\label{eq:Heisenberg}
H_0=-\sum_{{\bf R}, {\bf R}'}J({\bf R}-{\bf R}'){\bf S}({\bf R})\cdot{\bf S}({\bf R}')
-h\sum_{{\bf R}}S_z({\bf R})\ , 
\end{equation}
where $J({\bf R}-{\bf R}')$ is the exchange coupling between atoms
located at lattice sites $\bf R$ and $\bf R^\prime$, and $h$ is an
external magnetic field parallel to the $z$-axis. As proved by MW, in
the limit $h\to 0$ and $T\ne 0$ no spontaneous ferro- or
antiferromagnetic ordering occurs in 1D and 2D.

The DM interaction can be written as
\begin{equation}
\label{HDM}
H_{DM}=-2\sum_{{\bf R},{\bf R}'} {\bf D} ({\bf R}-{\bf R}')\cdot[{\bf S}({\bf
  R})\times {\bf S}({\bf R}')]\ , 
\end{equation}
where the antisymmetric Dzyaloshinskii-Moriya vector ${\bf D}$
satisfies ${\bf D}({\bf R}-{\bf R}')=-{\bf D}({\bf R}'-{\bf R})$.
Therefore, the total Hamiltonian of the system $H $ is 
\begin{equation}
\label{eq:totalHamiltonian}
  H = H_0 +H_ {DM} \ .
\end{equation}

\section{Method}

We first study the Heisenberg Hamiltonian and later on add
  the DM term.  A convenient procedure to tackle the problem at hand
is to use the Bogoliubov inequality in combination with the Fourier
transform of the spin ${\bf S}({\bf R})$. The latter is given by
\begin{equation}
  \label{eq:fourier}
  {\bf S}({\bf k}) = \sum_{\bf R} {\bf S}({\bf R}) e^{-i{\bf k \cdot R}} \ ; \qquad
  {\bf S}({\bf R}) = \frac 1N\sum_{\bf k} {\bf S}({\bf k}) e^{i{\bf k \cdot R}} \ ,
\end{equation}
where the ${\bf k}$ sum is limited to the first Brillouin
zone.

Let us now consider two operators, $A$ and $B$ whose statistical
average does exist.  Using the Schwartz inequality for the inner
product defined in reference~\cite{Bogoliubov62}, one obtains the
following Bogoliubov inequality:
\begin{equation}
\label{eq:AB}
\beta \langle \{A,A^{\dagger}\}\rangle\langle[[B,H],B^{\dagger}]\rangle\geq
 2|\langle [B,A]\rangle|^2 \ ,
\end{equation} 
where $\beta=1/(k_B T)$, $\{C,D \}= CD+DC$ and $[C,D] = CD-DC$.

Defining $A=S_-(-{\bf k}-{\bf K})$ and $B = S_+({\bf k})$ one obtains,
using Eq.~(\ref{eq:AB}), for the average spin per site that
\begin{eqnarray}
\label{HeisenbergSz}
&&\beta \sum_{{\bf k}}\langle \{ S_-(-{\bf k}-{\bf K}), S_+({\bf k}
+{\bf K}) \} \rangle\geq\nonumber\\  
&&\quad 2\sum_{{\bf k}} \frac{|\langle [ S_+({\bf k}),S_-(-{\bf k}-{\bf K})]\rangle|^2}
  {\langle[ [S_+({\bf k}),H_0],S_-(-{\bf k})]\rangle} \ .
\end{eqnarray}

The transverse components are obtained analogously, by defining
$A=S_-(-{\bf k}-{\bf K})$ and $B = S_z({\bf k})$, and using again the
Bogoliubov inequality. However, in contrast with the exchange
interaction, the DM term yields a spontaneous transverse
magnetization, at nonzero temperatures. 

Because of translational invariance
\begin{equation}
\sum_{{\bf k}}\langle S_i({\bf k})S_j(-{\bf k})\rangle = N^2  \langle
S_i({\bf R}_0)S_j({\bf R}_0)  \rangle \ , 
\end{equation}
and due to the symmetry of the exchange interaction $J(-{\bf
  k})=J({\bf k})$,  Eq.~(\ref{HeisenbergSz}) can be written as
\begin{equation}
\frac{\beta s(s+1)}{4s^2_z}\geq
\sum_{{\bf k}}\frac{1}{\Delta({\bf k})} \ ,
\label{HeisenbergSz2}
\end{equation}
where  $\Delta({\bf k})=\langle[ [S_+({\bf k}),H_0],S_-(-{\bf
  k})]\rangle $ and the spin per 
  particle is $s_z=|\langle S_z({\bf k})\rangle|/N$. 
In the continuum limit Eq.~(\ref{HeisenbergSz2}) takes the form
\begin{equation}
\frac{\rho \beta s(s+1)}{2s^2_z}\geq \left (\frac{k_0}{2\pi} \right )^D
\int\limits_{B.Z.} \frac{d{\bf u}} {\omega u^2+hs_z}\ ,
\label{eq:continuum}
\end{equation}
where $\omega=s(s+1)\sum_{{\bf
    R}} J({\bf R}) k_0^2R^2$, $1/\rho$ is the volume per spin, $D$ is
the dimensionality of the system, ${\bf u}={\bf k}/k_0$ is the normalized 
momentum and ${\bf k_0}$ is the first Brillouin vector. Since in the limit 
$h\to0$ for $D=1$ and $D=2$ the
right hand side of Eq.~(\ref{eq:continuum}) diverges, no spontaneous
magnetization occurs at finite temperatures.

\section{Results}
However, if the DM term is taken into consideration, and the full
Hamiltonian with finite range interactions given by
Eq.~(\ref{eq:totalHamiltonian}) is considered, then a spontaneous
magnetization does appear.  It just leads to the change $\Delta({\bf
  k})\to\Delta({\bf k})+\Delta_{DM}({\bf k})$, where
\begin{equation}
\label{DM}
\Delta_{DM}({\bf k})=\langle[[S_+({\bf k}),H_{DM}],S_-(-{\bf k})] \rangle \ ,
\end{equation}

On the other hand the spin components obey the following inequality:
\begin{equation}
\sum_{{\bf k}}\langle S_i({\bf k})S_j(-{\bf k})\rangle \leq N^2
\langle S_i({\bf R}_0)S_j({\bf R}_0)  \rangle \ ,
\end{equation}
which, combined with ${\bf D}(-{\bf k})={\bf D}({\bf k})$ and using 
{$D_{\pm}({\bf R})=D_y({\bf R})\pm i D_x({\bf R})$}, yields 
\begin{equation}
\label{eq:Delta-DM}
\Delta_{DM}({\bf k})\leq 2N\bigg(\frac {\gamma k}{k_0}
+ \gamma_0 \bigg) \ ,
\end{equation} 
where
\begin{eqnarray}
\gamma &=& 4s(s+1)\sum_{{\bf R}} |D_z({\bf R})|k_0R \ > 0\ , \nonumber \\
\gamma_0 &=& 2s(s+1)\sum_{{\bf R}} ( |D_z({\bf R})| + |D_+({\bf R})| )
\ > 0
\ .
\label{eq:def-gammas}
\end{eqnarray}
We notice that, due to the antisymmetry of the DM vector, 
the first term in Eq.~(\ref{eq:Delta-DM}) is of order $k{\bf=|k|}$, while the 
exchange term is of order $k^2$. This way we obtain, in the
continuum limit and with $h \to 0$, the expression
\begin{equation}
\frac{\rho\beta s(s+1)}{2s^2_z}\geq \left (\frac{k_0}{2\pi} \right )^D
\int \limits_{B.Z.} \frac{d{\bf u}} { \omega u^2 + \gamma u +\gamma_0}
> 0 \ .
\label{eq:continuumDM}
\end{equation}

If the exchange terms are ignored, which corresponds to the limit
$\omega \to 0$, in one dimension the integration yields

\begin{equation}
 \frac{\rho \beta s(s+1)}{2s_z^2} 
  \geq \frac{k_0} {\pi} \ \frac {\ln(1+\gamma/\gamma_0)}  {\gamma} > 0 \ .
\end{equation}
 Consequently, infrared divergences ($u\to0$) are removed by the DM
 terms. Moreover, in the $\lim {\gamma_0 \to 0} $, {\it i.e.}  when the
 DM interaction is switched off, the Mermin-Wagner result is
 recovered, since 

 \begin{equation}
   \frac{\rho \beta s(s+1)}{2s_z^2} 
   \geq \lim_{\gamma \to 0} \frac{k_0} {\pi} \    \left [ \frac
{\ln(1+\gamma/\gamma_0)}  {\gamma}  \right ] 
= \frac{k_0} {\pi \gamma_0}  \ ,
 \end{equation}
 and thus

 \begin{equation}
   \lim_{\gamma_0 \to 0} \frac{k_0} {\pi \gamma_0} \to \infty \ .
 \end{equation}

 However, when the DM interaction is switched on, {\it i.e.} for
 $\gamma > 0$ and $\gamma_0 > 0$, the magnitude of the spin per
 particle is finite, and has both an upper and a lower bound, even in
 the absence of the exchange interaction  

  \begin{equation}
  \frac{\pi \gamma \rho \beta s(s+1)}{2k_0\ln(1+\gamma/\gamma_0) } 
   \geq s_z^2 > 0 \ ,
 \end{equation}
 since $\rho, \beta$ and $s$ are positive.  

In two dimensions, in the limit $\omega \to 0$, we obtain

 \begin{equation}
 \frac{\rho \beta s(s+1)}{s^2_z}\geq 
 \frac{k_0^2}{\pi} \ \frac
 {\gamma-\gamma_0\ln(1+\gamma/\gamma_0)}{\gamma^2} \ >0 \ ,
 \end{equation}
 and by the same token 
 
 \begin{equation}
 \frac{\pi \gamma^2 \rho \beta
   s(s+1)}{k_0^2(\gamma-\gamma_0\ln(1+\gamma/\gamma_0))}\geq  s^2_z >0 \ ,
 \end{equation}
 and we confirmed that the function
 $\gamma-\gamma_0\ln(1+\gamma/\gamma_0)$ is always positive.

Consequently, the DM interaction generates spontaneous magnetic
order at nonzero temperature in one and two dimensions.

On the other hand, for the transverse components of the spin we have
$\tilde{\Delta}_{DM}({\bf k})=\langle [[S_z({\bf k}), H_{DM}],
S_z(-{\bf k})\rangle$.  Using the Bogoliubov inequality, in an
analogous way to Eq.~(\ref{eq:Delta-DM}), one obtains
$\tilde{\Delta}_{DM}({\bf k})\leq \omega_0$ with
\begin{equation}
\omega_0=4s(s+1)\sum_{{\bf R}}|D_+({\bf R})| \ .
\end{equation} 
Since in the continuum limit the contribution of this constant takes the 
form
 \begin{equation}
\frac{\rho\beta s(s+1)}{s^2_-}\geq \left( \frac{k_0}{2\pi} \right )^D
\int\limits_{B.Z.} \frac{d{\bf u}}
{ \omega u^2 + \omega_0} \ ,
\label{eq:DMs-}
\end{equation}
there is both an upper and a lower bound for the transverse component
$s_-^2$ in one- and two-dimensions, in the limit $\omega \to 0$, as well.

The DM term Eq.~(\ref{HDM}), which corresponds to the antisymmetric
part of the Moriya tensor \cite{MoriyaPRL60,MoriyaPR60}, and derived
from an extension of the Anderson theory of superexchange
\cite{Anderson50,Anderson59}, is linear in the spin-orbit
coupling. Hence it represents the leading-order contribution of this
interaction. However, as pointed out by Schekhtman {\it et al.}
\cite{Aharony92}, since the antisymmetric part of the DM interaction
$|{\bf D}({\bf R}-{\bf R}')|$ is of the order of $(\Delta g/g)J({\bf
  R}-{\bf R}')$ and the symmetric one $|{\bf \Gamma}({\bf R}-{\bf
  R}')|$ is of the order $(\Delta g/g)^2J({\bf R}-{\bf R}')$, with $g$
the gyromagnetic ratio and $\Delta g $ the deviation from the free
electron value, then
\begin{equation}
  |{\bf \Gamma}({\bf R}-{\bf R}')|\sim|{\bf D}({\bf R}-{\bf
    R}')|^2/J({\bf R}-{\bf R}')  \ .
\end{equation}
and therefore the contribution of the symmetric part also must be
taken into account. In addition, performing a local spin rotation in
the direction of the antisymmetric DM vector $\hat{{\bf d}}({\bf
  R}-{\bf R}')={\bf D}({\bf R}-{\bf R}')/|{\bf D}({\bf R}-{\bf R}')|$,
the antisymmetric and symmetric DM interactions can be reduced to an
isotropic exchange interaction as shown by Schekhtman et
    al. \cite{Aharony92}. In fact, let us consider an isotropic
Heisenberg model with symmetric and antisymmetric components of the DM
interaction. The Hamiltonian of this system is given by
\begin{eqnarray}
&H&=-\sum_{{\bf R}, {\bf R}'}\bigg[ J({\bf R}-{\bf R}'){\bf S} ({\bf
  R})\cdot{\bf S}({\bf R}')\nonumber\\ 
&+&2{\bf D}({\bf R}-{\bf R}')\cdot({\bf S}({\bf R})\times{\bf S}({\bf R}'))\nonumber\\
&+&\Gamma({\bf R}-{\bf R}')\bigg( 2(\hat{{\bf d}}\cdot{\bf S}({\bf
  R})) (\hat{{\bf d}}\cdot{\bf S}({\bf R}')) - {\bf S}({\bf R})\cdot{\bf S}({\bf R}')
\bigg) \bigg] \ . \nonumber\\
\end{eqnarray}
Using a local rotation transformation of the form 
\begin{eqnarray}
{\bf S}({\bf R})&=&(1-\cos\theta)(\hat{{\bf d}}\cdot\tilde{{\bf S}}({\bf R}))\hat{{\bf d}}
\nonumber\\
&&~~+\cos\theta\,\tilde{{\bf S}}({\bf R}) + \sin\theta\,\tilde{{\bf S}}({\bf R})\times\hat{{\bf d}},\\
{\bf S}({\bf R}')&=&(1-\cos\theta)(\hat{{\bf d}}\cdot\tilde{ {\bf S}}({\bf R}'))\hat{{\bf d}}
\nonumber\\
&&~~+\cos\theta\,\tilde{{\bf S}}({\bf R}') - \sin\theta\,\tilde{{\bf S}}({\bf R}')\times\hat{{\bf d}},
\label{eq:trans}
\end{eqnarray}
which correspond to a spin at site ${\bf R}({\bf R}')$ rotated around  the $\hat{{\bf d}}$ 
axis by the angles $\theta$ and $-\theta$, respectively, and where
$\sin \theta=D/\sqrt{J^2+D^2}$ and $\cos \theta=J/\sqrt{J^2+D^2}$, one obtains
\begin{equation}
H=-\sum_{{\bf R}, {\bf R}'}\bigg(J({\bf R}-{\bf R}')+\frac{D({\bf R}-{\bf R}')}{J({\bf R}-{\bf R}')} 
\bigg)\tilde{{\bf S}}({\bf R})\cdot\tilde{{\bf S}}({\bf R}').
\label{eq:totalH}
\end{equation}
where spin rotated variables preserve the commutation relation,
\begin{equation}
[\tilde{S}_i({\bf R}), \tilde{S}_j({\bf
  R}')]=i\epsilon_{ijk}\tilde{S}_k({\bf R})\delta({\bf R}-{\bf R}') \ .
\end{equation}

The MW theorem establishes that there is no spontaneous symmetry
breaking in one and two dimensional systems. This feature is preserved
by the DM interaction, since local rotation transformations could be
also be defined with opposite chirality ($\theta\to-\theta$), and
hence there is a local chiral invariance underlying the breakdown of
rotational symmetry invariance that precludes the spontaneous symmetry
breaking.  Moreover, Imry and Ma~\cite{Imry75} determined that domain
formation is energetically favorable against weak random fields, in
one and two dimensional systems. And Berezinskii~\cite{berezinskii70}
established that in classical systems long range fluctuations develop
finite values at large distances. In this sense the magnetic ordering
induced by the DM interaction seems to be more closely related to
topological ordering, induced some kind of Kosterlitz-Thouless
transition~\cite{Kosterlitz73}. Consequently, the elementary
excitations of the system cannot be expanded in terms of standard spin
waves.

The DM vector determines the direction of the local rotation of the
initial spin variables, and therefore it is possible to find a ground
state with a net spin average per particle at finite
temperature. However, in spite of the fact that interaction is
anisotropic there is no anisotropy energy induced by the DM
interaction. 

Finally, the terms contributed by the symmetric components of the DM
interaction, not included in the exchange interaction, have a constant
contribution to the denominators of Eqs.~(\ref{eq:continuumDM}) and
(\ref{eq:DMs-}), of the form
\begin{eqnarray}
\tilde{\omega}&=&s(s+1)\sum_{{\bf R}} (|\hat \Gamma({\bf R})| +
4|\Gamma_+({\bf R})|),\\ 
\tilde{\omega}_{0}&=&s(s+1)\sum_{{\bf R}} |\Gamma_{+}({\bf R})|,
\end{eqnarray}
where
\begin{eqnarray}
\Gamma_+({\bf R})&=&4\Gamma({\bf R})\hat{{\bf
    d}}_z\hat{{\bf d}}_+\ , \\
\hat \Gamma({\bf R})&=&\Gamma({\bf R})(\hat{{\bf
    d}}^2_+-\hat{{\bf d}}^2_-) \ , \\
\hat{{\bf d}}_{\pm}&=&\hat{{\bf d}}_x\pm i \hat{\bf d}_y \ .
\end{eqnarray}
Since these terms lift the infrared divergences they induce a net spin per 
particle in one and two dimensions at finite temperatures. 

\section{Conclusion}
In summary, using the Bogoliuvov inequality we have demonstrated that
a short range antisymmetric or symmetric interaction, as the DM one,
generates a long range ordering in low dimensional systems.  In spite
of the fact that in the ground state there are no coherent long range
spin waves, locally the DM interaction induces a small correlation
between the spins that reduce quantum fluctuations and induce a net
average spin per particle at finite temperature.

In view of the fact that the Mermin-Wagner theorem constitutes an
important landmark in the understanding of ordering of magnetic
systems which are adequately described by the Heisenberg Hamiltonian,
it is significant that the introduction of the Dzyaloshinsky-Moriya
interaction \cite{Dzyaloshinsky58,MoriyaPRL60,MoriyaPR60} allows for
the stabilization of magnetic order \cite{Crepieux98} in one and two
dimensions at finite temperatures.  Furthermore, it also contributes
to the understanding of the two dimensional order in layered cuprates,
as already pointed out by Kastner et al.~\cite{Kastner98}, and also to
understand the stabilization mechanism of two dimensional layers and
membranes where long-wavelength fluctuations destroy long-range order.

In closing we underline that i)~the DM interaction does not
  break rotational invariance; ii)~the short range order DM
  interaction does supress quantum fluctuations in low dimensional
  systems; and, iii)~the particle permutation symmetry of the DM
  interaction yields contributions to the denominator of
  Eq.~(\ref{eq:continuumDM}), that are of order $k^{m}$ with $m<2$,
  which remove divergences and thus allow for the existence of long
  range magnetic order in one and two dimensions.

\acknowledgments
This work was supported by the \textit{Fondo Nacional de
  Investigaciones Cien\-t{\'\i}\-fi\-cas y Tecnol{\'o}gicas}
(FONDECYT, Chile) under grants \#1120356 (DA), 1090225 and 1120399 (MK),
and by the {\it Financiamiento Basal para Centros Cient\'\i ficos y
  Tecnol\'ogicos de Excelencia}, CEDENNA Project. We also acknowledge
the support by FIC-MINECOM FB10-061F



\end{document}